\documentclass[12pt]{article}
\begin{document}
\title{The nonlinear time-dependent response of isotactic polypropylene}

\author{Aleksey D. Drozdov and Jesper deClaville Christiansen\\
Department of Production\\
Aalborg University\\
Fibigerstraede 16\\
DK--9220 Aalborg, Denmark}
\date{}
\maketitle

\begin{abstract}
Tensile creep tests, tensile relaxation tests and a tensile
test with a constant rate of strain are performed on injection-molded
isotactic polypropylene at room temperature
in the vicinity of the yield point.
A constitutive model is derived for the time-dependent
behavior of semi-crystalline polymers.
A polymer is treated as an equivalent network of chains bridged
by permanent junctions.
The network is modelled as an ensemble of passive meso-regions 
(with affine nodes) and active meso-domains
(where junctions slip with respect to their positions in the bulk medium
with various rates).
The distribution of activation energies for sliding in active meso-regions 
is described by a random energy model.
Adjustable parameters in the stress--strain relations are found by
fitting experimental data.
It is demonstrated that the concentration of active meso-domains
monotonically grows with strain, whereas the average potential
energy for sliding of junctions and the standard deviation of
activation energies suffer substantial drops at the yield point.
With reference to the concept of dual population of crystalline lamellae,
these changes in material parameters are attributed to
transition from breakage of subsidiary (thin) lamellae in the sub-yield region
to fragmentation of primary (thick) lamellae in the post-yield region of deformation.
\end{abstract}
\newpage

\section{Introduction}
This paper is concerned with the experimental study and modelling 
of the time-dependent behavior of isotactic polypro\-py\-lene (iPP) at strains up to 20 \%
in uniaxial tensile tests, creep tests and relaxation tests
at room temperature.
Isotactic polypropylene is chosen for the investigation because of
its numerous applications in industry 
(oriented films for packaging,
reinforcing fibres,
nonwoven fabrics,
blends with thermoplastic elastomers, etc.).

The nonlinear viscoelastic response of polypropylene was studied
by Ward and Wolfe (1966) and Smart and Williams (1972) three decades ago,
and, in recent years, 
by Wortmann and Schulz (1994, 1995),
Ariyama (1996),
Ariyama et al. (1997),
Dutta and Edward (1997),
Read and Tomlins (1997)
Tomlins and Read (1998),
and Sweeney et al. (1999).

Viscoplasticity and yielding of iPP have been investigated 
in the past five years by Kalay and Bevis (1997),
Coulon et al. (1998),
Seguela et al. (1999),
Staniek et al. (1999),
Nitta and Takayanagi (1999, 2000)
and Labour et al. (2001),
to mention a few.

Dynamic mechanical analysis reveals that the loss tangent 
of iPP demonstrates two pronounced maxima 
being plotted versus temperature (Andreassen, 1999;
Seguela et al., 1999; Lopez-Manchado and Arroyo, 2000).
The first maximum ($\beta$--transition in the interval 
between $T=-20$ and $T=10$ $^{\circ}$C) is associated with 
the glass transition in a mobile part of the amorphous phase, 
whereas the other maximum ($\alpha$--transition in the interval 
between $T=70$ and $T=110$ $^{\circ}$C) is attributed to the
glass transition in the remaining part of the amorphous phase,
the so-called ``rigid amorphous fraction" (Verma et al., 1996).
This conclusion is confirmed by DSC (differential scanning calorimetry)
traces of quenched polypropylene
which show (in the heating mode) an endotherm near $T=70$ $^{\circ}$C 
ascribed to thermal activation of amorphous domains with 
restricted mobility (Seguela et al., 1999).

Isotactic polypropylene is a semi-crystalline polymer containing
three different crystallographic forms:
monoclinic $\alpha$ crystallites,
hexagonal $\beta$ structures,
orthorhombic $\gamma$ polymorphs,
together with ``smectic" mesophase (Iijima and Strobl, 2000).
At rapid cooling of the melt (at the stage of injection molding),
$\alpha$ crystallites and smectic mesophase are mainly developed,
whereas $\beta$ and $\gamma$ polymorphs are observed as minority components
(Kalay and Bevis, 1997; Al-Raheil et al., 1998) that disappear 
after annealing above 130 $^{\circ}$C (Al-Raheil et al., 1998; Labour et al., 2001).

A unique feature of $\alpha$ spherulites in iPP is the 
lamellar cross-hatching: 
development of transverse lamellae oriented in the direction 
perpendicular to the direction of radial lamellae 
(Iijima and Strobl, 2000; Maiti et al., 2000).
The characteristic size of $\alpha$ spherulites in injection-molded 
specimens is estimated as 100 to 200 $\mu$m 
(Kalay and Bevis, 1997; Coulon et al., 1998).
These spherulites consist of crystalline lamellae with thickness 
of 10 to 20 nm (Coulon et al., 1998; Maiti et al., 2000).
The amorphous phase is located  (i) between spherulites,
(ii) inside spherulites, in ``liquid pockets" (Verma et al., 1996)
between lamellar stacks,
and (iii) between lamellae in lamellar stacks.
It consists of (i) relatively mobile chains between spherulites, in liquid pockets
and between radial lamellae inside lamellar stacks,
and (ii) severely restricted chains in the regions bounded 
by radial and tangential lamellae.

Stretching of iPP specimens results in inter-lamellar separation,
rotation and twist of lamellae,
fine and coarse slip of lamellar blocks and
their fragmentation (Aboulfaraj et al., 1995; Seguela et al., 1999),
chain slip through the crystals, 
sliding and breakage of tie chains (Nitta and Takayanagi, 1999, 2000),
and activation of rigid amorphous fraction.
At large strains, these morphological transformations lead to
cavitation, formation of fibrills and stress-induced crystallization
(Zhang et al. 1999).

It is hard to believe that these mechanically-induced changes in the micro-structure of iPP
can be adequately described by a constitutive model 
with a small number of adjustable parameters.
To develop stress--strain relations, we apply a method of ``homogenization of micro-structure,"
according to which a sophisticated morphology of isotactic polypropylene
is modelled by an equivalent phase whose deformation captures essential 
features of the response of this semi-crystalline polymer.
We choose a network of chains as the equivalent phase for the following reasons:
\begin{enumerate}
\item
The viscoelastic response of isotactic polypropylene
is conventionally associated with rearrangement of chains 
in amorphous regions (Coulon et al., 1998).

\item
Sliding of tie chains along and their detachment from lamellae 
play the key role in the time-dependent response of iPP
(Nitta and Takayanagi, 1999, 2000).

\item
The viscoplastic flow in semi-crystalline polymers
is assumed to be ``initiated in the amorphous phase 
before transitioning into the crystalline phase" 
(Meyer and Pruitt, 2001).

\item
The time-dependent behavior of polypropylene is conventionally modelled
within the concept of a network of macromolecules
(Sweeney and Ward, 1995, 1996; Sweeney et al., 1999).
\end{enumerate}

Isotactic polypropylene at room temperature (i.e., above the glass transition 
temperature for the mobile amorphous phase) is treated a permanent network 
of macromolecules bridged by junctions (physical cross-links,
entanglements and lamellar blocks).
The network is assumed to be highly heterogeneous
(this inhomogeneity is attributed to interactions between lamellae
and surrounding amorphous regions, as well as to local density fluctuations
in the amorphous phase),
and it is thought of as an ensemble of meso-regions (MR)
with different potential energies.
Two types of MRs are distinguished: 
(i) active domains, where junctions can slide with respect to their
positions in the bulk material as they are thermally agitated
(the mobile part of the amorphous phase), 
and (ii) passive domains, where sliding of junctions is prevented 
by surrounding lamellae (the rigid amorphous fraction).

Straining of a specimen induces 
\begin{enumerate}
\item
Sliding of meso-domains with respect to
each other (which reflects fragmentation and coarse slip of lamellae).

\item
Sliding of junctions with respect to their positions in the 
stress-free medium (which is associated with slip of tie molecules 
along lamellae and fine slip of lamellar blocks).
\end{enumerate}
Sliding of MRs with respect to each other is modelled as a rate-independent
process that describes the elastoplastic response of iPP.
Sliding of junctions in meso-domains is treated as a
rate-dependent phenomenon that reflects the viscoplastic behavior of
isotactic polypropylene.

Stretching of a specimen results in an increase in the concentration
of active MRs and changes in the distribution of meso-domains with
various activation energies for sliding of junctions driven by release 
of part of the rigid amorphous fraction due to lamellar fragmentation.

The objective of this study is two-fold:
\begin{enumerate}
\item
To report experimental data in a tensile test with a constant strain
rate, in creep tests and in relaxation tests at several elongation ratios
on injection-molded iPP specimens annealed for 24 h at the temperature 
$T=140$ $^{\circ}$C.

\item
To derive constitutive equations for the time-dependent response
of a semi-crystalli\-ne polymer and to find adjustable parameters
in the stress--strain relations by fitting observations.
\end{enumerate}
In previous studies on modelling the response
of amorphous and semi-crystalline polymers in the sub-yield and
post-yield regions,
see, e.g., pioneering works by Haward and Thackray (1968) and
G'Sell and Jonas (1979), and more recent publications by
Boyce et al. (1988),
Bordonaro and Krempl (1992),
Arruda et al. (1995),
Hasan and Boyce (1995),
Krempl and Bordonaro (1995, 1998),
Zhang and Moore (1997),
Spathis and Kontou (1998),
Drozdov (2001),
Duan et al. (2001),
stress--strain curves, creep curves and relaxation curves
were treated independently of each other (in the sence that
different adjustable parameters were determined by matching observations
in different tests).
The aim of the present paper is to approximate experimental data
in three conventional types of mechanical tests within one constitutive model.
This allows two approaches in the nonlinear viscoelasticity with 
an ``internal time" to be compared, the so-called models with 
stress- and strain-induced material clocks 
(Lustig et al., 1996; Krempl and Bordonaro, 1998,
Wineman, 2002), as well as to shed some light
on a mechanism for mechanically-induced changes in 
relaxation (retardation) spectra in the vicinity of the yield point.

The exposition is organized as follows.
The specimens and the experimental procedure are described
in Section 2.
The distribution of meso-regions with various potential energies
for sliding of junctions is introduced in Section 3.
Kinetic equations for sliding of MRs with respect to each other are
proposed in Section 4.
Sliding of junctions in active meso-domains is modelled in Section 5.
The strain energy density of a semi-crystalline polymer
is determined in Section 6.
Constitutive equations are derived in Section 7
by using the laws of thermodynamics.
These equations are further simplified to describe observations
in ``rapid" tensile tests, creep tests and relaxation tests.
Section 8 is concerned with fitting experimental data.
Our findings are briefly discussed in Section 9.
Some concluding remarks are formulated in Section 10.

\section{Experimental procedure}

Isotactic polypropylene (Novolen 1100L) was supplied by BASF (Targor).
ASTM dumbbell specimens were injection molded 
with length 148 mm, width 10 mm and height 3.8 mm.
To erase the influence of thermal history, the samples were annealed 
in an oven at 140 $^{\circ}$C for 24 h and slowly cooled by air.
To minimize the effect of physical aging on the time-dependent response of iPP,
mechanical tests were carried out a week after the thermal pre-treatment.

Uniaxial tensile relaxation tests were performed 
at room temperature on a testing machine Instron--5568 
equipped with electro-mechanical sensors 
for the control of longitudinal strains in the active zone of samples 
(with the distance 50 mm between clips).
The tensile force was measured by a standard load cell.
The engineering stress, $\sigma$, was determined
as the ratio of the axial force to the cross-sectional area
of the specimens in the stress-free state.
The specimens were loaded with the cross-head speed 5 mm/min 
(that corresponded to the Hencky strain rate 
$\dot{\epsilon}_{H}=1.1\cdot 10^{-3}$ s$^{-1}$), which provides
nearly isothermal test conditions (Arruda et al., 1995).

The engineering stress, $\sigma$, is plotted versus the elongation ratio
$\lambda$ in Figure 1.
The true longitudinal stress, $\sigma_{\rm t}$, is calculated as
$\sigma_{\rm t}=\sigma\lambda$ (this formula is based 
on the incompressibility condition).
The true stress is also depicted in Figure 1, which shows
that necking of samples does not occur at elongations up to 50 \%.
The apparent yield strain, $\epsilon_{\rm y1}$, calculated as the 
intersection point of tangent lines to the true stress--elongation ratio
diagram at ``small" and ``large" deformations, equals 0.04.
The yield strain, $\epsilon_{\rm y2}$, determined as the strain corresponding
to the maximum on the engineering stress--engineering strain curve,
equals 0.08.

A series of 8 creep tests was performed at the engineering stresses
$\sigma_{1}^{0}=10.0$ MPa,
$\sigma_{2}^{0}=15.0$ MPa,
$\sigma_{3}^{0}=20.0$ MPa,
$\sigma_{4}^{0}=25.0$ MPa,
$\sigma_{5}^{0}=30.0$ MPa,
$\sigma_{6}^{0}=30.38$ MPa,
$\sigma_{7}^{0}=30.94$ MPa, and
$\sigma_{8}^{0}=32.80$ MPa.
The last three values of stress correspond to the initial strains 
(at the beginning of the creep tests) 
$\epsilon_{6}^{0}=0.45$,
$\epsilon_{7}^{0}=0.50$, and
$\epsilon_{8}^{0}=0.60$.
The first four tests were carried out in the sub-yield region of deformation
(the initial strains are less than the yield strain $\epsilon_{\rm y1}$),
whereas the other tests were performed in the interval 
between the yield strains $\epsilon_{\rm y1}$ and $\epsilon_{\rm y2}$.

Each creep test was carried out on a new sample.
In the $m$th test ($m=1,\ldots,8$), a specimen was loaded 
with the cross-head speed 5 mm/min up to the engineering stress $\sigma_{m}^{0}$
that was preserved constant during the creep test, $t_{\rm c}=20$ min.
The longitudinal strains, $\epsilon$, measured in the first 6 tests 
are plotted versus  the logarithm ($\log=\log_{10}$) of time $t$ 
(the instant $t=0$ corresponds to the beginning of a creep test) in Figure 2.
This figure demonstrates that the rate of increase in strain, $\epsilon$,
with time, $t$,
is relatively low at small stresses, and it noticeably grows with $\sigma$.
The creep curves at stresses $\sigma_{m}^{0}$ exceeding 20 MPa are
plotted in Figure 3, where 
the Hencky strain $\epsilon_{\rm H}=\ln \lambda$ is presented as a function
of time $t$.
Figure 3 reveals the primary creep of iPP at the stress $\sigma_{4}^{0}$,
transition from the primary creep to the secondary creep at the stress $\sigma_{5}^{0}$,
and transition from the secondary creep to the ternary creep at higher stresses,
$\sigma_{6}^{0}$ to $\sigma_{8}^{0}$.
These transitions are indicated by lines AA$^{\prime}$ ($\epsilon_{\rm H}=0.06$) 
and BB$^{\prime}$ ($\epsilon_{\rm H}=0.14$) in Figure 3.

A series of 4 relaxation tests was performed at the strains
$\epsilon_{1}^{0}=0.05$,
$\epsilon_{2}^{0}=0.10$,
$\epsilon_{3}^{0}=0.15$, and
$\epsilon_{4}^{0}=0.20$.
The first test was carried out at the strain belonging to the interval 
between the yield strains $\epsilon_{\rm y1}$ and $\epsilon_{\rm y2}$, 
whereas the other tests were performed at strains in the post-yield
region of deformation.

Any relaxation test was carried out on a new sample.
In the $m$th test ($m=1,\ldots,4$), a specimen was loaded 
with the cross-head speed 5 mm/min up to the longitudinal strain $\epsilon_{m}^{0}$
that was preserved constant during the relaxation time $t_{\rm r}=20$ min.
The engineering stress, $\sigma$, is plotted versus  the logarithm of time $t$ 
(the instant $t=0$ corresponds to the beginning of a relaxation test) in Figure 4.
This figure shows that the relaxation curves are strongly
affected by strain.

\section{Distribution of meso-regions}
 
A semi-crystalline polymer is treated as a permanent network of 
chains bridged by junctions.
The network is thought of as an ensemble of meso-regions with
various potential energies for slippage of junctions with respect
to their positions in the reference state.
Two types of meso-domains are distinguished: passive and active.
In passive MRs, all nodes are assumed to move affinelly
with the bulk material.
In active MRs, the junctions slide with respect to their
reference positions under loading.

Sliding of junctions in active MRs is modelled as
a thermally activated process.
The rate of sliding in a MR with potential energy $\bar{\omega}$ is given 
by the Eyring equation (Eyring, 1936)
\[
\Gamma=\Gamma_{0}\exp\biggl (-\frac{\bar{\omega}}{k_{\rm B}T}\biggr ),
\]
where $k_{\rm B}$ is Boltzmann's constant, 
$T$ is the absolute temperature, 
and the pre-factor $\Gamma_{0}$ is independent
of energy $\bar{\omega}$ and temperature $T$.
Confining ourselves to isothermal processes and introducing 
the dimensionless activation energy $\omega=\bar{\omega}/(k_{\rm B}T_{0})$, 
where $T_{0}$ is some reference temperature,
we arrive at the formula
\begin{equation}
\Gamma(\omega) =\Gamma_{0}\exp (-\omega).
\end{equation}
Some lamellae (that restrict mobility of junctions in passive MRs) break under straining,
which implies an increase in the concentration of active meso-domains.
As a result, the number of strands in active MRs grows,
and the number of strands in passive meso-domains decreases.
Denote by  $N_{\rm a}(t,\omega)$ the number of strands (per unit mass)
in active meso-domains with energy $\omega$ at instant $t\geq 0$.
The total number of strands in active MRs, $X_{\rm a}(t)$, reads
\begin{equation}
X_{\rm a}(t)=\int_{0}^{\infty} N_{\rm a}(t,\omega) d\omega .
\end{equation}
The number of strands (per unit mass) in passive MRs, $X_{\rm p}(t)$,
is connected with the number of strands in active meso-domains, $X_{\rm a}(t)$,
by the conservation law
\begin{equation}
X_{\rm a}(t) +X_{\rm p}(t) =X,
\end{equation}
where $X$ is the number of active strands per unit mass
(this quantity is assumed to be time-independent).

The distribution of strands in active MRs is described by the ratio, $p(t,\omega)$, 
of the number, $N_{\rm a}(t,\omega)$, of strands in active meso-domains 
with energy $\omega$ to the total number of strands in active MRs,
\begin{equation}
p(t,\omega)=\frac{N_{\rm a}(t,\omega)}{X_{\rm a}(t)},
\end{equation}
and by the concentration, $\kappa(t)$, of active MRs,
\begin{equation}
\kappa(t)=\frac{X_{\rm a}(t)}{X}.
\end{equation}
In what follows, constitutive equations for a semi-crystalline polymer will be derived
for an arbitrary distribution of active MRs.
To fit experimental data, the random energy model  is employed with
\begin{equation}
p(t,\omega) = p_{0}(t) \exp \biggl [ -\frac{(\omega-\Omega(t))^{2}}{2\Sigma^{2}(t)} \biggr ]
\quad (\omega\geq 0), \qquad
p(t,\omega)=0 \quad (\omega <0).
\end{equation}
Here $\Omega$ is the average activation energy in an ensemble of
active meso-domains,
$\Sigma$ is the standard deviation of potential energies for sliding
of junctions,
and $p_{0}$ is determined by the condition
\begin{equation}
\int_{0}^{\infty} p(t,\omega) d\omega =1.
\end{equation}

\section{Sliding of meso-regions}

It is assumed that meso-domains are not rigidly connected,
but can slide with respect to each other under straining.
Sliding of meso-domains is treated as a rate-independent process
and is associated with the elastoplastic behavior of a semi-crystalline
polymer.
We suppose that an increase in strain, $\epsilon$, by an increment,
$d\epsilon$, causes growth of the elastoplastic strain, $\epsilon_{\rm ep}$,
by an increment, $d\epsilon_{\rm ep}$, that is proportional to $d\epsilon$,
\[
 d\epsilon_{\rm ep}=\varphi d\epsilon.
\]
The coefficient of proportionality, $\varphi$, may, in general, be a function of
strain, $\epsilon$, stress, $\sigma$, and the elastoplastic strain, $\epsilon_{\rm ep}$.
Only the dependence of $\varphi$ on $\epsilon$
is taken into account, which results in the kinematic equation
\begin{equation}
\frac{d\epsilon_{\rm ep}}{dt}(t)=\varphi (\epsilon(t) ) \frac{d\epsilon}{dt}(t),
\qquad
\epsilon_{\rm ep}(0)=0.
\end{equation}
The function $\varphi(\epsilon)$ vanishes at $\epsilon=0$ 
(the elastoplastic strain equals zero at very small strains),
monotonically increases with strain,
and reaches some limiting value $b\in (0,1)$ at rather large strains
(which corresponds to a steady regime of plastic flow).
To reduce the number of adjustable parameters in the constitutive equations,
an exponential dependence is adopted,
\begin{equation}
\varphi(\epsilon)=b\Bigl [ 1-\exp(-a\epsilon)\Bigr ],
\end{equation}
where the positive coefficients $a$ and $b$ are found by matching observations.

Equations (8) and (9) differ from conventional flow rules in elastoplasticity,
where the elastoplastic strain is assumed to be proportional to the stress, $\sigma$.
Similarities and differences between our approach and traditional ones
will be discussed in Section 8.

\section{Sliding of junctions in active MRs}

Sliding of junctions in active meso-domains with respect to their
positions in a stress-free medium is treated as a rate-dependent
process and is associated with the viscoelastic response of a semi-crystalline
polymer.
Sliding of junctions in active MRs reflects (i) sliding of tie chains along lamellae,
(ii) slip of chains in amorphous regions with respect to entanglements, 
and (iii) rearrangement of junctions driven by slip of lamellar blocks.

The non-affine deformation of a network is modelled as a mechanically
activated process induced by straining of active meso-domains.
The strain in a meso-region, $e$, is defined as the difference between 
the macro-strain, $\epsilon$, and the elastoplastic strain, $\epsilon_{\rm ep}$,
caused by sliding of meso-domains with respect to each other,
\[
e(t)=\epsilon(t)-\epsilon_{\rm ep}(t). 
\]
Accepting the first-order kinetics for sliding of junctions, 
\[
\frac{\partial \epsilon_{\rm ve}}{\partial t}(t,\omega)
=\Gamma(\omega)\Bigl [ e(t)-\epsilon_{\rm ve}(t,\omega)\Bigr ],
\]
and using Eq. (1), we arrive at the constitutive equation
\begin{equation}
\frac{\partial \epsilon_{\rm ve}}{\partial t}(t,\omega)
=\Gamma_{0}\exp(-\omega)\Bigl [ \epsilon(t)-\epsilon_{\rm ep}(t)-\epsilon_{\rm ve}(t,\omega)\Bigr ],
\qquad
\epsilon_{\rm ve}(0,\omega)=0,
\end{equation}
where $\epsilon_{\rm ve}(t,\omega)$ is the strain driven by sliding of junctions
in an active MR with potential energy $\omega$.

\section{Strain energy density}

The elastic strain, $\epsilon_{\rm e}$, is calculated as the difference between
the macro-strain, $\epsilon$, and the strains, $\epsilon_{\rm ep}$ and
$\epsilon_{\rm ve}$, induced by sliding of meso-domains with respect to
each other and by sliding of junctions in active MRs with respect to
their reference positions.

For a passive meso-region (where sliding of junctions is prevented),
the elastic strain is given by
\[ 
\epsilon_{\rm e}(t)=\epsilon(t)-\epsilon_{\rm ep}(t). 
\]
For an active meso-domain with potential energy $\omega$, the elastic
strain reads
\[ 
\epsilon_{\rm e}(t,\omega)=\epsilon(t)-\epsilon_{\rm ep}(t)-\epsilon_{\rm ve}(t,\omega). 
\]
A strand is modelled as a linear elastic medium with the strain energy
\begin{equation}
w=\frac{1}{2}\mu \epsilon_{\rm e}^{2},
\end{equation}
where $\mu$ is a constant rigidity.

Multiplying the mechanical energy of a strand, Eq. (11), 
by the number of strands per unit mass,
summing the strain energies for strands in active and passive meso-domains,
and neglecting the energy of inter-chain interaction, we find the mechanical
energy per unit mass of a semi-crystalline polymer,
\begin{equation}
W(t)=\frac{1}{2} \mu \biggl [ \int_{0}^{\infty} N_{\rm a}(t,\omega) \Bigl (\epsilon(t)-
\epsilon_{\rm ep}(t)-\epsilon_{\rm ve}(t,\omega)\Bigr )^{2}d\omega
+X_{\rm p}(t)\Bigl (\epsilon(t)-\epsilon_{\rm ep}(t)\Bigr )^{2} \biggr ].
\end{equation}
To develop a stress--strain relation, an explicit expression is necessary
for the derivative of $W$ with respect to time.
Differentiation of Eq. (12) implies that
\begin{eqnarray}
\frac{dW}{dt}(t) &=& \mu \biggl [ \int_{0}^{\infty} N_{\rm a}(t,\omega)
\Bigl (\epsilon(t)-\epsilon_{\rm ep}(t)-\epsilon_{\rm ve}(t,\omega)\Bigr )d\omega
+X_{\rm p}(t) \Bigl (\epsilon(t)-\epsilon_{\rm ep}(t) \Bigr )\biggr ]
\nonumber\\
&& \times \Bigl [\frac{d\epsilon}{dt}(t)-\frac{d\epsilon_{\rm ep}}{dt}(t)\Bigr ]
-Y_{1}(t)-Y_{2}(t),
\end{eqnarray}
where
\begin{eqnarray}
Y_{1}(t) &=& -\frac{1}{2}\mu \biggl [ \int_{0}^{\infty} \frac{\partial N_{\rm a}}{\partial t}(t,\omega)
\Bigl (\epsilon(t)-\epsilon_{\rm ep}(t)-\epsilon_{\rm ve}(t,\omega)\Bigr )^{2} d\omega
\nonumber\\
&& +\frac{d X_{\rm p}}{dt}(t) \Bigl (\epsilon(t)-\epsilon_{\rm ep}(t) \Bigr )^{2} \biggr ],
\\
Y_{2}(t) &=& \mu \int_{0}^{\infty} N_{\rm a}(t,\omega)
\Bigl (\epsilon(t)-\epsilon_{\rm ep}(t)-\epsilon_{\rm ve}(t,\omega)\Bigr )
\frac{\partial \epsilon_{\rm ve}}{\partial t}(t,\omega) d\omega.
\end{eqnarray}
It follows from Eqs. (2), (3), (8) and (13) that
\begin{eqnarray}
\frac{dW}{dt}(t)  &=& \mu \biggl [ X \Bigl (\epsilon(t)-\epsilon_{\rm ep}(t) \Bigr )
- \int_{0}^{\infty} N_{\rm a}(t,\omega) \epsilon_{\rm ve}(t,\omega) d\omega \biggr ]
\Bigl [ 1-\varphi(\epsilon(t))\Bigr ]\frac{d\epsilon}{dt}(t)
\nonumber\\
&&-Y_{1}(t)-Y_{2}(t).
\end{eqnarray}
Equations (2) and (3) imply that
\[ \frac{dX_{\rm p}}{dt}(t)=-\frac{dX_{\rm a}}{dt}(t)=-\int_{0}^{\infty} \frac{\partial N_{\rm a}}{\partial t}(t,\omega)
d\omega.
\]
This equality together with Eq. (14) results in
\begin{equation}
Y_{1}(t)=\frac{1}{2}\mu  \int_{0}^{\infty} \frac{\partial N_{\rm a}}{\partial t}(t,\omega)
\biggl [ \Bigl (\epsilon(t)-\epsilon_{\rm ep}(t) \Bigr )^{2}
-\Bigl (\epsilon(t)-\epsilon_{\rm ep}(t)-\epsilon_{\rm ve}(t,\omega)\Bigr )^{2}
\biggr ] d\omega.
\end{equation} 
Combining Eqs. (10) and (15), we find that
\begin{equation}
Y_{2}(t) = \mu \int_{0}^{\infty} N_{\rm a}(t,\omega)\Gamma(\omega)
\Bigl (\epsilon(t)-\epsilon_{\rm ep}(t)-\epsilon_{\rm ve}(t,\omega)\Bigr )^{2} d\omega.
\end{equation}

\section{Constitutive equation}

For isothermal uniaxial deformation, the Clausius--Duhem inequality reads
\begin{equation}
-\frac{dW}{dt}+\sigma\frac{d\epsilon}{dt}\geq 0.
\end{equation}
Substituting expression (16) into Eq. (19) and using Eqs. (4) and (5), we obtain
\begin{eqnarray}
&& \biggl [ \sigma(t)-E\Bigl (1-\varphi(\epsilon(t))\Bigr )
\Bigl (\epsilon(t)-\epsilon_{\rm ep}(t)-\kappa(t)\int_{0}^{\infty} p(t,\omega)\epsilon_{\rm ve}(t,\omega)d\omega
\Bigr )\biggr ] \frac{d\epsilon}{dt}(t)
\nonumber\\
&& +Y_{1}(t)+Y_{2}(t)\geq 0,
\end{eqnarray}
where $E=\mu X$.

It follows from Eqs. (17) and (18) that for an active loading program
(when $\epsilon(t)$, $\epsilon_{\rm ve}(t)$, $\epsilon_{\rm ep}(t)$ 
and $N_{\rm a}(t,\omega)$ increase with time),
the functions $Y_{1}(t)$ and $Y_{2}(t)$ are non-negative.
This means that the dissipation inequality (20) is satisfied, provided that
the expression in the square brackets vanishes, 
which results in the stress--strain relation
\begin{equation}
\sigma(t) =E\Bigl (1-\varphi(\epsilon(t))\Bigr ) \biggl [\epsilon(t)-\epsilon_{\rm ep}(t)
-\kappa(t)\int_{0}^{\infty} p(t,\omega)\epsilon_{\rm ve}(t,\omega)d\omega \biggr ].
\end{equation}
Given functions $\kappa(t)$, $\Omega(t)$ and $\Sigma(t)$, constitutive equations
(6), (8) to (10) and (21) describe the time-dependent response of a semi-crystalline
polymer at uniaxial deformation.
These relations are determined by 4 adjustable parameters: an analog of Young's modulus
$E$ in Eq. (21), dimensionless constants $a$ and $b$ in Eq. (9),
and the attempt rate for sliding of junctions $\Gamma_{0}$ in Eq. (10).
The pre-factor $\Gamma_{0}$ can be excluded from the governing equations
by introducing a ``shifted" potential energy $\tilde{\omega}=\omega-\omega_{0}$
with $\omega_{0}=\ln \Gamma_{0}$.
Because this transformation does not change the structure of the stress--strain relations,
we set $\Gamma_{0}=1$ s without loss of generality.

For ``rapid" deformations, when the effect of sliding of junctions in active MRs
on the mechanical response of a specimen is negligible,
the constitutive equations are substantially simplified.
Neglecting the integral term in Eq. (21) and using Eq. (9), we find that
\begin{equation}
\sigma =E\Bigl [1-b\Bigl ( 1-\exp(-a\epsilon)\Bigr ) \Bigr ] (\epsilon-\epsilon_{\rm ep}),
\end{equation}
where the elastoplastic strain $\epsilon_{\rm ep}$ obeys Eqs. (8) and (9),
\begin{equation}
\frac{d\epsilon_{\rm ep}}{{d\epsilon}}=b\Bigl ( 1-\exp(-a\epsilon)\Bigr ),
\qquad
\epsilon_{\rm ep}(0)=0.
\end{equation}
Equations (22) and (23) are determined by 3 material constants, $E$, $a$ and $b$,
to be found by matching a stress--strain curve for a tensile test with a constant
strain rate.

To study ``slow" deformation processes, when sliding of junctions in meso-domains
is to be taken into account, an additional hypothesis should be introduced to describe
the effect of deformation history on the quantities $\kappa$, $\Omega$ and $\Sigma$.
Two approaches are conventionally used to predict the effect of mechanical 
factors on these parameters, the so-called models with strain and stress clocks.
For a survey of these concepts, the reader is referred to Drozdov (1998) 
and the bibliography therein.

The theory of a stress-induced internal time is traditionally employed to fit observations
in creep tests with the loading program
\begin{equation}
\sigma(t)=\left \{\begin{array}{ll}
0, & t<0,\\
\sigma^{0}, & t\geq 0,
\end{array} \right .
\end{equation}
where $\sigma^{0}$ is a given stress.
In terms of our model, this concept means that the quantities $\kappa$, $\Omega$ and $\Sigma$ 
depend on the current stress $\sigma$.
Combining Eqs. (6), (8), (9), (21) and (24), we arrive at the formulas
\begin{eqnarray}
\epsilon(t) &=& \epsilon_{\rm ep}(t)+p^{0}\kappa(\sigma^{0})\int_{0}^{\infty}
\epsilon_{\rm ve}(t,\omega)\exp\Bigl [-\frac{(\omega-\Omega(\sigma^{0}))^{2}}{2\Sigma^{2}(\sigma^{0})}
\Bigr ]d\omega 
\nonumber\\
&&+(\epsilon^{0}-\epsilon_{\rm ep}^{0}) 
\frac{1-b(1-\exp(-a\epsilon^{0}))}{1-b(1-\exp(-a\epsilon(t)))},
\\
\frac{d\epsilon_{\rm ep}}{dt}(t) &=& b\Bigl ( 1-\exp(-a\epsilon(t))\Bigr )
\frac{d\epsilon}{dt}(t),
\qquad
\epsilon_{\rm ep}(0)=\epsilon_{\rm ep}^{0}.
\end{eqnarray}
In these equations, the initial instant $t=0$ corresponds to the beginning of
the creep test, 
the quantities $\epsilon^{0}$ and $\epsilon_{\rm ep}^{0}$
are found by integration of Eqs. (22) and (23) in the interval from $\sigma=0$ 
to $\sigma=\sigma^{0}$,
the coefficient $p^{0}$ is determined by condition (7),
and the function $\epsilon_{\rm ve}(t,\omega)$ obeys Eq. (10).

It is worth noting that Eq. (26) can be integrated explicitly to express the elastoplastic
strain, $\epsilon_{\rm ep}$, by means of the macro-strain $\epsilon$.
However, we do not dwell on this transformation.

Given a stress, $\sigma^{0}$, Eqs. (10), (25) and (26) are determined by 3 experimental
constants, $\kappa(\sigma^{0})$, $\Omega(\sigma^{0})$ and $\Sigma (\sigma^{0})$ to be
found by matching observations in a creep test.

According to the concept of a strain-induced material clock, the parameters
$\kappa$, $\Omega$ and $\Sigma$ are functions of the current strain $\epsilon$.
This approach is conventionally used to approximate experimental data
in a relaxation test with
\begin{equation}
\epsilon(t)=\left \{\begin{array}{ll}
0, & t<0,\\
\epsilon^{0}, & t\geq 0,
\end{array} \right .
\end{equation}
where $\epsilon^{0}$ is a given strain.
It follows from Eqs. (8), (9) and (27) that the elastoplastic strain, $\epsilon_{\rm ep}$,
is time-independent. 
The quantity $\epsilon_{\rm ep}(t)=\epsilon_{\rm ep}^{0}$ is determined by integration
of Eq. (23) from zero to $\epsilon^{0}$.
Combining Eqs. (6), (21) and (27), we find that
\begin{eqnarray}
\sigma(t) &=& E\Bigl [ 1-b\Bigl (1-\exp(-a\epsilon^{0})\Bigr )\Bigr ]
\Bigl \{ \epsilon^{0}-\epsilon_{\rm ep}^{0}
\nonumber\\
&&-p^{0}\kappa(\epsilon^{0})\int_{0}^{\infty} \epsilon_{\rm ve}(t,\omega)
\exp\Bigl [-\frac{(\omega-\Omega(\epsilon^{0}))^{2}}{2\Sigma^{2}(\epsilon^{0})} \Bigr ]d\omega \Bigr \},
\end{eqnarray}
where the instant $t=0$ corresponds to the beginning of a relaxation test,
the coefficient $p^{0}$ is determined by condition (7),
and the function $\epsilon_{\rm ve}(t,\omega)$ is governed by Eq. (10)
with $\Gamma_{0}=1$,
\[
\frac{\partial \epsilon_{\rm ve}}{\partial t}(t,\omega)
=\exp(-\omega)\Bigl [ \epsilon^{0}-\epsilon_{\rm ep}^{0}-\epsilon_{\rm ve}(t,\omega)\Bigr ],
\qquad
\epsilon_{\rm ve}(0,\omega)=0.
\]
Introducing the notation
\begin{equation}
e_{\rm ve}(t,\omega)=\frac{\epsilon_{\rm ve}(t,\omega)}{\epsilon^{0}-\epsilon_{\rm ep}^{0}},
\end{equation}
we present the latter equation as follows:
\begin{equation}
\frac{\partial e_{\rm ve}}{\partial t}(t,\omega)
=\exp(-\omega)[ 1-e_{\rm ve}(t,\omega) ],
\qquad
e_{\rm ve}(0,\omega)=0.
\end{equation}
Substitution of expression (29) into Eq. (28) results in
\begin{equation}
\sigma(t) = \sigma^{0}(\epsilon^{0}) \Bigl \{ 1-p^{0}\kappa(\epsilon^{0})\int_{0}^{\infty} e_{\rm ve}(t,\omega)
\exp\Bigl [-\frac{(\omega-\Omega(\epsilon^{0}))^{2}}{2\Sigma^{2}(\epsilon^{0})} \Bigr ]d\omega \Bigr \},
\end{equation}
where
\[
\sigma^{0}(\epsilon^{0})=E\Bigl [ 1-b\Bigl (1-\exp(-a\epsilon^{0})\Bigr )\Bigr ]
(\epsilon^{0}-\epsilon_{\rm ep}^{0}).
\]
Given a strain, $\epsilon^{0}$, Eqs. (30) and (31) are characterized by 4 experimental
constants, $\sigma^{0}(\epsilon^{0})$, $\kappa(\epsilon^{0})$, $\Omega(\epsilon^{0})$ 
and $\Sigma (\epsilon^{0})$ to be determined by fitting observations in a relaxation test.

Our aim now is to find adjustable parameters in the constitutive equations by
matching experimental data depicted in Figures 1, 2 and 4, and to assess the
applicability of the concept of internal time with stress- and strain-induced material
clocks.

\section{Fitting of observations}

We begin with the approximation of the stress--strain curve depicted
in Figure 5.
The restriction on strains ($\epsilon_{\max}=0.06$) may explained by
two reasons: (i) at higher elongations, the assumption that the strain energy
of a strand is a quadratic function of strain, see Eq. (11), becomes questionable,
and (ii) according to Figure 3, at $\epsilon=0.06$ the primary creep 
is transformed into the secondary creep (developed plastic flow), which
is beyond the scope of the present study.

Under uniaxial tension with the cross-head speed 5 mm/min, the strain $\epsilon_{\max}=0.06$
is reached within 69 s. 
According to Figure 2, changes in strain induced by sliding of junctions during this
period are insignificant at stresses up to 20 MPa, whereas
the duration of stretching at higher stresses does not exceeds 30 s,
which causes rather small growth of strains.
Based on these observations, we treat the deformation process as rapid 
and apply Eqs. (22) and (23) to match observations.
To find the constants $E$, $a$ and $b$, we fix the intervals 
$[0,a_{\max}]$ and $[0,b_{\max}]$, 
where the ``best-fit" parameters $a$ and $b$ are assumed to be located,
and divide these intervals into $J$ subintervals by
the points $a_{i}=i\Delta a$ and $b_{j}=j\Delta b$  ($i,j=1,\ldots,J$)
with $\Delta a=a_{\max}/J$ and $\Delta b=b_{\max}/J$.
For any pair, $\{ a_{i}, b_{j} \}$, we integrate Eqs. (22) and (23)
numerically (with the step $\Delta \epsilon=5.0\cdot 10^{-5}$)
by the Runge--Kutta method.
The elastic modulus $E=E_{0}(i,j)$ is found by the least-squares 
method from the condition of minimum of the function
\[
K(i,j)=\sum_{\epsilon_{l}} \Bigl [ \sigma_{\rm exp}(\epsilon_{l})
-\sigma_{\rm num}(\epsilon_{l}) \Bigr ]^{2},
\]
where the sum is calculated over all experimental points, $\epsilon_{l}$,
depicted in Figure 5, $\sigma_{\rm exp}$ is the engineering stress 
measured in the tensile test, 
and $\sigma_{\rm num}$ is given by Eq. (22).
The ``best-fit" parameters $a$ and $b$ minimize 
$K$ on the set $ \{ a_{i}, b_{j} \quad (i,j=1,\ldots, J)  \}$.
After determining their values, $a_{i}$ and $b_{j}$, 
this procedure is repeated twice for the new intervals $[ a_{i-1}, a_{i+1}]$
and $[ b_{j-1}, b_{j+1}]$ to ensure an acceptable accuracy of fitting.
The ``best-fit" parameters read $E= 2.12$ GPa, $a=   38.10$
and $b=0.64$.
This value of $E$ slightly exceeds Young's modulus ($E=1.50$ GPa)
provided by the supplier for a virgin material, which may be explained 
by changes in the microstructure of spherulites at annealing.

To estimate the elastoplastic strain, $\epsilon_{\rm ep}$, and
the difference between the present model and the conventional flow rule 
\begin{equation}
\frac{d\epsilon_{\rm ep}}{d\epsilon}=k\sigma
\end{equation}
with a constant coefficient $k$,
we present results of numerical simulation in Figure 6.
This figure shows that the elastoplastic strain, $\epsilon_{\rm ep}$, 
is negligible at strains $\epsilon<0.025$, and it increases (practically linearly) 
with macro-strain at higher elongation ratios.
The ratio
\[
r(\epsilon)=\frac{\varphi(\epsilon)}{\sigma(\epsilon)}
\]
linearly grows with strain at $\epsilon < 0.025$, and it becomes practically
constant at $\epsilon > 0.03$.
The latter implies that in the region, where the influence of elastoplastic strains 
on the response of iPP cannot be disregarded, Eqs. (22) and (23) are rather close
to the flow rule (32).
An advantage of Eqs. (22) and (23) is that (i) they have a transparent physical meaning,
(ii) no additional constants (analogs of the yield stress, $\sigma_{\rm y}$,
or yield strain, $\epsilon_{\rm y}$) are to be introduced in the stress--strain relations, 
and (iii) these equations can correctly predict the time-dependent response of iPP
in creep tests (see later), where Eq. (32) appears to be an oversimplified relationship.

We proceed with fitting observations in creep tests depicted in Figure 2.
For any stress, $\sigma^{0}$, the quantities $\epsilon^{0}$ and $\epsilon_{\rm ep}^{0}$
are found by integration of Eqs. (22) and (23) with the material constants found in the
approximation of the stress--strain curve plotted in Figure 5.
The quantities $\kappa(\sigma^{0})$, $\Omega(\sigma^{0})$ and $\Sigma (\sigma^{0})$
are determined by the following algorithm.
We fix the intervals $[0,\kappa_{\max}]$, $[0,\Omega_{\max}]$ and $[0,\Sigma_{\max}]$, 
where the ``best-fit" parameters $\kappa$, $\Omega$ and $\Sigma$ are assumed to be located,
and divide these intervals into $J$ subintervals by
the points $\kappa_{i}=i\Delta \kappa$, $\Omega_{j}=j \Delta \Omega$ 
and $\Sigma_{k}=k\Delta \Sigma$ ($i,j,k=1,\ldots,J$)
with $\Delta \kappa=\kappa_{\max}/J$, $\Delta \Omega=\Omega_{\max}/J$
and $\Delta \Sigma=\Sigma_{\max}/J$.
For any pair, $\{ \Omega_{j}, \Sigma_{k} \}$, the constant $p^{0}=p^{0}(j,k)$
is determined by Eq. (7), where the integral is evaluated 
by Simpson's method with 200 points and the step $\Delta \omega=0.15$.
For any triple, $\{ \kappa_{i}, \Omega_{j}, \Sigma_{k} \}$, 
Eqs. (10), (25) and (26) are integrated numerically 
(with the time step $\Delta t=0.1$) by the Runge--Kutta method.
The ``best-fit" parameters $\kappa$, $\Omega$ and $\Sigma$ 
are found from the condition of minimum of the function
\[
K(i,j,k)=\sum_{t_{l}} \Bigl [ \epsilon_{\rm exp}(t_{l})-\epsilon_{\rm num}(t_{l}) \Bigr ]^{2},
\]
where the sum is calculated over all experimental points, $t_{l}$,
presented in Figure 2, 
$\epsilon_{\rm exp}$ is the strain measured in the creep test, 
and $\epsilon_{\rm num}$ is given by Eq. (25).
After determining the ``best-fit" values, $\kappa_{i}$, $\Omega_{j}$ and $\Sigma_{k}$, 
this procedure is repeated for the new intervals $[ \kappa_{i-1}, \kappa_{i+1}]$,
$[ \Omega_{j-1}, \Omega_{j+1}]$ and $[ \Sigma_{k-1}, \Sigma_{k+1}]$ 
to provide an acceptable accuracy of fitting.
Figure 2 demonstrates fair agreement between the experimental data and
the results of numerical simulation.

The adjustable parameters $\Omega$, $\Sigma$ and $\kappa$ are plotted versus
the engineering stress, $\sigma$, in Figures 7 to 9.
The experimental data are approximated by the linear functions
\begin{equation}
\Omega=\Omega_{0}+\Omega_{1}\sigma,
\qquad
\Sigma=\Sigma_{0}+\Sigma_{1}\sigma,
\qquad
\kappa=\kappa_{0}+\kappa_{1}\sigma,
\end{equation}
where the coefficients $\Omega_{m}$, $\Sigma_{m}$ and $\kappa_{m}$ ($m=0,1$)
are found by the least-squares technique.
Figures 7 and 8 show that the quantities $\Omega$ and $\Sigma$ increase with
stress, $\sigma$, reach their maxima in the interval between $\sigma=25$ and $\sigma=30$ MPa
(which means, in the vicinity of the yield strain $\epsilon_{\rm y1}$),
and dramatically decrease at higher stresses.
According to Figure 9, the concentration of active meso-regions, $\kappa$, 
increases with strain up to $\sigma=25$ MPa, whereas at higher strains, 
the slope of the straightline (33) noticeably grows.

Finally, we approximate the experimental data in the relaxation tests
presented in Figure 4.
To fit observations, we re-write Eq. (31) as
\begin{equation}
\sigma(t) = C_{1}(\epsilon^{0})+p^{0}C_{2}(\epsilon^{0})\int_{0}^{\infty} e_{\rm ve}(t,\omega)
\exp\Bigl [-\frac{(\omega-\Omega(\epsilon^{0}))^{2}}{2\Sigma^{2}(\epsilon^{0})} \Bigr ]d\omega 
\end{equation}
with
\[
C_{1}=\sigma^{0},
\qquad
C_{2}=-\sigma^{0}\kappa.
\]
For any strain, $\epsilon^{0}$, we fix the intervals $[0,\Omega_{\max}]$ and $[0,\Sigma_{\max}]$, 
where the ``best-fit" parameters $\Omega$ and $\Sigma$ are located,
and divide these intervals into $J$ subintervals by
the points $\Omega_{i}=i \Delta \Omega$ and $\Sigma_{j}=j\Delta \Sigma$ ($i,j=1,\ldots,J$)
with $\Delta \Omega=\Omega_{\max}/J$ and $\Delta \Sigma=\Sigma_{\max}/J$.
For any pair, $\{ \Omega_{i}, \Sigma_{j} \}$, the constant $p^{0}=p^{0}(i,j)$
is determined from Eq. (7), where the integral is evaluated 
by Simpson's method with 200 points and the step $\Delta \omega=0.15$.
Equations (30) are integrated numerically (with the time step $\Delta t=0.1$) 
by the Runge--Kutta method.
The constants $C_{1}$ and $C_{2}$ in Eq. (34) are determined by the least-squares
algorithm to minimize the function
\[
K(i,j)=\sum_{t_{l}} \Bigl [ \sigma_{\rm exp}(t_{l})-\sigma_{\rm num}(t_{l}) \Bigr ]^{2},
\]
where the sum is calculated over all experimental points, $t_{l}$,
presented in Figure 4, 
$\sigma_{\rm exp}$ is the stress measured in the relaxation test, 
and $\sigma_{\rm num}$ is given by Eq. (34).
The ``best-fit" parameters $\Omega$ and $\Sigma$ are found 
from the condition of minimum of the function $K(i,j)$.
After determining the ``best-fit" values, $\Omega_{i}$ and $\Sigma_{j}$, 
this procedure is repeated for the new intervals 
$[ \Omega_{i-1}, \Omega_{i+1}]$ and $[ \Sigma_{j-1}, \Sigma_{j+1}]$ 
to guarantee good accuracy of fitting.
Figure 4 demonstrates an acceptable agreement between the experimental data 
and the results of numerical analysis.

The quantities $\Omega$, $\Sigma$ and $\kappa=-C_{2}/C_{1}$ are plotted versus
strain $\epsilon$ in Figures 10 to 12.
The experimental data are approximated by the linear functions
\begin{equation}
\Omega=\Omega_{0}+\Omega_{1}\epsilon,
\qquad
\Sigma=\Sigma_{0}+\Sigma_{1}\epsilon,
\qquad
\kappa=\kappa_{0}+\kappa_{1}\epsilon,
\end{equation}
where the coefficients $\Omega_{m}$, $\Sigma_{m}$ and $\kappa_{m}$ ($m=0,1$)
are found by the least-squares technique.
Figures 10 to 12 demonstrate that the adjustable parameters $\Omega$, $\Sigma$ and $\kappa$
found by matching observations in relaxation tests 
monotonically increase with strain $\epsilon$.

There is no doubt that Eqs. (30) and (31) can be applied to fit experimental data
in the relaxation test with the smallest strain $\epsilon_{1}^{0}=0.05$.
Some questions arise, however, whether these equations (based on 
the quadratic approximation (11) of the strain energy), may be used 
to match observations in relaxation tests at higher strains.
Fortunately, it can be shown that Eqs. (30) and (31) remain valid for an arbitrary
(not necessary quadratic in strains) mechanical energy per strand $w$.
To avoid complicated algebra employed in the derivation of constitutive equations
at finite strains, appropriate transformations are omitted.

\section{Discussion}

Our objective now is to compare adjustable parameters in the constitutive 
equations found by fitting observations in the creep and relaxation tests.

First, we plot the quantities $\Omega$, $\Sigma$ and $\kappa$ determined 
in the approximation of the relaxation curve at $\epsilon_{1}^{0}$ together 
with the data obtained by matching the creep curves (Figures 7 to 9).
For this purpose, we replace the strain in the relaxation test, $\epsilon_{1}^{0}$, 
by the corresponding stress at the beginning of the relaxation process,
$\sigma^{0}=C_{1}(\epsilon_{1}^{0})$.
Figures 7 and 8 demonstrate good agreement between the parameters $\Omega$
and $\Sigma$ determined by fitting the experimental data in the creep and relaxation tests.
Figure 9 reveals that the concentration of active meso-domains, $\kappa$, found in
the relaxation text follows Eq. (33) with the coefficients
found in the approximation of the creep curves below the yield point, $\epsilon_{\rm y1}$,
see curve 1.
These conclusions confirm changes in the distribution function $p$ in the interval
$[\epsilon_{\rm y1}, \epsilon_{\rm y2}]$ revealed by matching the experimental
data in the creep tests, but question appropriate findings for the concentration
of active MRs $\kappa$.

To shed some light on this controversy, we re-plot the experimental constants
$\Omega$, $\Sigma$ and $\kappa$ found by fitting observations in the creep tests
below the yield strain $\epsilon_{\rm y1}$ together with those determined 
in the approximation of the relaxation curves (Figures 10 to 12).
The adjustable parameters $\Omega$, $\Sigma$ and $\kappa$ found by matching
the creep curves are depicted as functions of strain at the beginning
of creep tests.
Figures 10 and 11 reveal that the quantities $\Omega$ and $\Sigma$ linearly
increase with strain up to the yield strain $\epsilon_{\rm y1}$, where they
suffer a pronounced drop, and proceed to grow with strain afterwards.
Figure 12 demonstrates that the concentration of active meso-domains, $\kappa$,
linearly increases with strain in the entire region of strains under consideration,
and the data found in the creep tests are in fair agreement with those determined
by matching the relaxation curves.

Based on these observations and adopting the dual lamellar population model
(Verma et al., 1996), we propose the following scenario for the effect 
of mechanical factors on the distribution of active MRs:
\begin{enumerate}
\item
Below the yield strain $\epsilon_{\rm y1}$, the average activation energy, $\Omega$,
and the standard deviation of activation energies, $\Sigma$, for sliding of junctions
in active MRs monotonically increase with strain $\epsilon$.
This growth is attributed to fragmentation of thin (subsidiary) lamellae formed
during injection-molding of specimens and developed at annealing of iPP.
The increase in the average activation energy, $\Omega$, is associated 
with 
\begin{itemize}
\item
breakage of subsidiary and transverse lamellae into small pieces 
that serve as extra physical cross-links in amorphous regions,

\item
mechanically-induced activation of rigid amorphous fraction,
where sliding of junctions was prevented by surrounding lamellae
in stress-free specimens.
\end{itemize}
The increase in the standard deviation of potential energies, $\Sigma$, reflects
heterogeneity of the fragmentation process, which grows with strain
because of the inhomogeneity of breakage of subsidiary lamellae 
in active meso-regions with different sizes (potential energies).

\item
When the strain, $\epsilon$, reaches the yield strain, $\epsilon_{\rm y1}$,
subsidiary (thin) lamellae become totally disintegrated, which results in 
a decrease in the average activation energy $\Omega$ (lamellar blocks that 
served as physical cross-links disappear) and a pronounced decrease 
in the standard deviation of potential energies $\Sigma$ (meso-domains become 
more homogeneous).
Figures 7 and 8 reveal that the interval of strains, where this ``homogenization"
of the micro-structure occurs, is relatively narrow (less than 2 \%), 
which implies that the slopes of curves 2 in these figures are rather high.

\item
Straining of a specimen above the yield strain, $\epsilon_{\rm y1}$,
causes fragmentation of dominant (thick) lamellae, which results in
a noticeable increase in $\Omega$ and $\Sigma$.
This growth is driven by the same mechanism as an increase in
$\Omega$ and $\Sigma$ in the sub-yield region of deformation
(breakage of lamellae into small pieces that serve as extra physical
cross-links in the amorphous phase).
This conclusion may be confirmed by the similarity of slopes 
of curves 1 and 2 for the standard deviation of potential energies of MRs,
$\Sigma$, depicted in Figure 11. 
The slopes of the curves 1 and 2 for the average potential energy for sliding
of junctions, $\Omega$, in Figure 10 differ substantially.
This discrepancy is attributed to the fact that the growth of the average
potential energy of active meso-domains in the sub-yield region of deformation
is governed by two morphological transformations (release of rigid amorphous fraction 
and formation of extra junctions in active MRs), 
whereas only the latter process takes place in the post-yield region.

\item
The concentration of active meso-domains, $\kappa$, monotonically increases
under stretching.
A noticeable increase in $\kappa$ with stress reported in Figure 9 is an artifact
caused by the assumption that the concentration of active MRs remains
constant during a creep test (which, in turn, is based on a concept of stress-induced
internal time).
Despite good agreement between the experimental data and the
results of numerical simulation revealed in Figure 2,
it seems more adequate to presume that the adjustable parameters 
$\Omega$, $\Sigma$ and $\kappa$ depend on the current strain 
(not stress).
This hypothesis does not affect curves 1 to 4 in Figure 2
(because of relatively small increases in strain during the creep
tests), but it results in substantial changes in creep curves 5 and 6
(corresponding to the yield region).
Assuming $\kappa$ to be a function of strain,
one can treat the experimental points on curve 2 in Figure 9
as some ``average" (over the creep curves) values of the concentration 
of active MRs in the post-yield region of deformation.
In this case, the values of $\kappa$ found by matching creep curves 5 and 6
in Figure 2 become quite comparable with the data depicted in Figure 12,
where these values roughly correspond to the vicinity of the yield point $\epsilon_{\rm y2}$.
\end{enumerate}

\section{Concluding remarks}

A model has been developed for the elastoplastic and viscoplastic
responses of semi-crystalline polymers at isothermal uniaxial
deformation with small strains.
To derive constitutive equations, a complicated micro-structure of a polymer
is replaced by an equivalent permanent network of macromolecules 
bridged by junctions (physical cross-links, entanglements and lamellae).
The network is thought of as an ensemble of meso-regions with various 
potential energies for sliding of junctions with respect to their positions
in a stress-free medium.

The elastoplastic (rate-independent) behavior of a semi-crystalline polymer 
is attribut\-ed to sliding of meso-domains with respect to each other.
The kinetics of sliding is governed by Eqs. (8) and (9) with 
2 adjustable parameters: $a$ and $b$.
The viscoplastic (rate-dependent) response is associated with sliding
of junctions in active MRs.
The sliding process is assumed to be thermally-agitated, and its rate
is described by the Eyring equation (1).
Equation (10) for the rate of viscoplastic strain is based on
the first order kinetics of sliding, and it does not contain experimental
constants (the attempt rate $\Gamma_{0}$ is set to be 1 s).

The distribution of active meso-domains is determined by the random 
energy model, Eq. (6), with three adjustable parameters:
the average activation energy for sliding of nodes, $\Omega$,
the standard deviation of potential energies of active MRs, $\Sigma$,
and the concentration of active meso-domains, $\kappa$,
that are affected by mechanical factors.
With reference to the concept of mechanically-induced material clocks,
two hypotheses are analyzed: (i) the quantities $\Omega$, $\Sigma$
and $\kappa$ are stress-dependent, and (ii) these parameters
are governed by the current strain.

Stress--strain relations are derived by using the laws of thermodynamics.
The constitutive equations are developed under the assumptions that
(i) a strand can be modelled as a linear elastic medium with the strain
energy (11), and (ii) the energy of inter-chain interaction is negligible.
These equations contain 6 material constants: an elastic modulus $E$,
and the quantities $a$, $b$, $\Omega$, $\Sigma$ and $\kappa$.

To find these parameters, a tensile test with a constant strain rate,
a series of 8 creep tests, and a series of 4 relaxation tests have been
performed on injection-molded isotactic polypropylene at room temperature.
Material constants are determined by fitting experimental data.
Figures 2, 4 and 5 demonstrate fair agreement between the observations
and the results of numerical simulation.

The following conclusions are drawn:
\begin{enumerate}
\item
Stretching of iPP in the sub-yield region of deformation
increases the average potential energy for sliding of junctions 
and the standard deviation of potential energies of active MRs.
This increase is attributed to fragmentation of subsidiary lamellae
and activation of the rigid amorphous fraction.

\item
In the vicinity of the yield point, the parameters $\Omega$ and $\Sigma$
fall down noticeably, which is associated with disappearance of
physical cross-links formed by blocks of disintegrated thin lamellae.

\item
In the post-yield region of deformation, the quantities $\Omega$ and $\Sigma$
grow with strain, which is explained by fragmentation of
dominant lamellae.

\item
The concentration of active meso-regions, $\kappa$, monotonically increases
with strain both in the sub-yield and post-yield domains.

\item
The hypothesis about a strain-induced material clock appears to be more
adequate to morphological transformations in isotactic polypropylene
under loading than the assumption about the dependence of material
parameters on the current stress.
\end{enumerate}
This study focused on modelling the nonlinear elastoplastic and viscoplastic
behavior of iPP in creep and relaxation tests at relatively small strains.
Some important questions remained, however, beyond the scope of
the present work.
In particular, transitions from the primary creep to the secondary and
tertiary creeps have not been analyzed.
No attention has been paid to the effect of time and temperature of annealing 
on the time-dependent response of isotactic polypropylene.
The applicability of the constitutive equations to the description of the mechanical
behavior of other semi-crystalline polymers has not been examined.
These issues will be the subject of a subsequent publication.
\newpage
\section*{References}
\parindent 0 mm

Aboulfaraj, M., C. G'Sell, B. Ulrich, and A. Dahoun,
``{\it In situ} observation of the plastic deformation of
polypropylene spherulites under uniaxial tension and simple
shear in the scanning electron microscope,"
Polymer {\bf 36}, 731--742 (1995).

Al-Raheil, I.A., A.M. Qudah, M. Al-Share,
``Isotactic polypropylene crystallized from the melt.
2. Thermal melting behavior,"
J. Appl. Polym. Sci. {\bf 67}, 1267--1271 (1998).

Andreassen, E.,
``Stress relaxation of polypropylene fibres with various
morphologies,"
Polymer {\bf 40}, 3909--3918 (1999).

Ariyama, T.,
``Viscoelastic-plastic behaviour with mean strain 
changes in polypropylene,"
J. Mater. Sci. {\bf 31}, 4127--4131 (1996).

Ariyama, T., Y. Mori, and K. Kaneko,
``Tensile properties and stress relaxation of polypro\-pylene 
at elevated temperatures,"
Polym. Eng. Sci. {\bf 37}, 81--90 (1997).

Arruda, E.M., M.C. Boyce, and R. Jayachandran, 
``Effects of strain rate, temperature and thermomechanical coupling 
on the finite strain deformation of glassy polymers,"
Mech. Mater. {\bf 19}, 193--212 (1995).

Bordonaro, C.M. and E. Krempl,  
``The effect of strain rate on the deformation and relaxation behavior
of 6/6 nylon at room temperature,"
Polym. Eng. Sci. {\bf 32}, 1066--1072 (1992).

Boyce, M.C., D.M. Parks, and A.S. Argon, 
``Large inelastic deformation of glassy polymers.
1. Rate dependent constitutive model,"
Mech. Mater. {\bf 7}, 15--33 (1988).

Coulon, G., G. Castelein, and C. G'Sell, 
``Scanning force microscopic investigation of plasticity
and damage mechanisms in polypropylene spherulites under
simple shear,"
Polymer {\bf 40}, 95--110 (1998).

Drozdov, A.D.,
``Mechanics of Viscoelastic Solids,"
John Wiley \& Sons, Chichester (1998).

Drozdov, A.D.,
``A model for the viscoelastic and viscoplastic responses of glassy polymers,"
Int. J. Solids Structures {\bf 38}, 8285--8304 (2001).

Duan, Y., A. Saigal, R. Grief, and M.A. Zimmerman,
``A uniform phenomenological constitutive model for glassy and
semicrystalline polymers,"
Polym. Eng. Sci. {\bf 41}, 1322--1328 (2001).

Dutta, N.K. and G.H. Edward, 
``Generic relaxation spectra of solid polymers. 
1. Development of spectral distribution model 
and its application to stress relaxation of polypropylene,"
J. Appl. Polym. Sci. {\bf 66}, 1101--1115 (1997).

Eyring, H.,
``Viscosity, plasticity, and diffusion as examples of absolute reaction rates,"
J. Chem. Phys. {\bf 4}, 283--291(1936).

G'Sell, C. and J.J. Jonas, 
``Determination of the plastic behavior of solid polymers at constant
true strain rate,"
J. Mater. Sci. {\bf 14}, 583--591 (1979).

Hasan, O.A. and M.C. Boyce, 
``A constitutive model for the nonlinear viscoelastic viscoplastic
behavior of glassy polymers,"
Polym. Eng. Sci. {\bf 35}, 331--344 (1995).

Haward, R.N. and G. Thackray, 
``The use of a mathematical model to describe stress--strain
curves in glassy thermoplastics,"
Proc. Roy. Soc. London {\bf A302}, 453--472 (1968).

Iijima, M. and G. Strobl,
``Isothermal crystallization and melting of isotactic polypropylene
analyzed by time- and temperature-dependent small-angle X-ray
scattering experiments,"
Macromolecules {\bf 33}, 5204--5214 (2000).

Kalay, G. and M.J. Bevis,
``Processing and physical property relationships in injection-molded 
isotactic polypropylene. 
1. Mechanical properties,"
J. Polym. Sci. B: Polym. Phys. {\bf 35}, 241--263 (1997).

Krempl, E. and C.M. Bordonaro,
``A state variable model for high strength polymers,"
Polym. Eng. Sci. {\bf 35}, 310--316 (1995).

Krempl, E. and C.M. Bordonaro, 
``Non-proportional loading of nylon 66 at room temperature,"
Int. J. Plasticity {\bf 14}, 245--258 (1998).

Labour, T., C. Gauthier, R. Seguela, G. Vigier, Y. Bomal, and G. Orange,
``Influence of the $\beta$ crystalline phase on the mechanical
properties of unfilled and CaCO$_{3}$-filled polypropylene.
1. Structural and mechanical characterization,"
Polymer {\bf 42}, 7127--7135 (2001).

Lopez-Manchado, M.A. and M. Arroyo,
``Thermal and dynamic mechanical properties of polypropylene and short
organic fiber composites,"
Polymer {\bf 41}, 7761--7767 (2000).

Lustig, S.R., R.M. Shay, and J.M. Caruthers,
``Thermodynamic constitutive equations for materials with memory on a material
time scale,"
J. Rheol. {\bf 40}, 69--106 (1996).

Maiti, P., M. Hikosaka, K. Yamada, A. Toda, and F. Gu,
``Lamellar thickening in isotactic polypropylene with high tacticity
crystallized at high temperature,"
Macromolecules {\bf 33}, 9069--9075 (2000).

Meyer, R.W. and L.A. Pruitt, 
``The effect of cyclic true strain on the morphology, structure,
and relaxation behavior of ultra high molecular weight 
polyethylene,"
Polymer {\bf 42}, 5293--5306 (2001).

Nitta, K.-H. and M. Takayanagi, 
``Role of tie molecules in the yielding deformation
of isotactic polyprolylene,"
J. Polym. Sci. B: Polym. Phys. {\bf 37}, 357--368 (1999).

Nitta, K.-H. and M. Takayanagi, 
``Tensile yield of isotactic polypropylene in
terms of a lamellar-cluster model,"
J. Polym. Sci. B: Polym. Phys. {\bf 38}, 1037--1044 (2000).

Read, B.E. and P.E. Tomlins,
``Time-dependent deformation of polypropylene in response 
to different stress histories,"
Polymer {\bf 38}, 4617--4628 (1997).

Seguela, R., E. Staniek, B. Escaig, and B. Fillon,
``Plastic deformation of polypropylene in relation to
crystalline structure,"
J. Appl. Polym. Sci. {\bf 71}, 1873--1885 (1999).

Smart, J. and J.G. Williams, 
``A comparison of single-integral non-linear viscoelasticity
theories,"
J. Mech. Phys. Solids {\bf 20}, 313--324 (1972).

Spathis, G. and E. Kontou, 
``Experimental and theoretical description of the plastic
behaviour of semicrystalline polymers,"
Polymer {\bf 39}, 135--142 (1998).

Staniek, E., R. Seguela, B. Escaig, and P. Francois,
``Plastic behavior of monoclinic polypropylene under hydrostatic
pressure in compressive testing,"
J. Appl. Polym. Sci. {\bf 72}, 1241--1247 (1999).

Sweeney, J., T.L.D. Collins, P.D. Coates, and R.A. Duckett, 
``High temperature large strain viscoelastic behaviour of polypropylene
modeled using an inhomogeneously strained network,"
J. Appl. Polym. Sci. {\bf 72}, 563--575 (1999).

Sweeney, J. and I.M. Ward, 
``The modelling of multiaxial necking in polypropylene
using a sliplink-crosslink theory,"
J. Rheol. {\bf 39}, 861--872 (1995).

Sweeney, J. and I.M. Ward, 
``A constitutive law for large deformations of polymers at
high temperatures,"
J. Mech. Phys. Solids {\bf 44}, 1033--1049 (1996).

Tomlins, P.E. and B.E. Read,
``Creep and physical ageing of polypropylene:
a comparison of models,"
Polymer {\bf 39}, 355--367 (1998).

Verma, R., H. Marand, and B. Hsiao,
``Morphological changes during secondary crystallization and
subsequent melting in poly(ether ether ketone) as studied
by real time small angle X-ray scattering,"
Macromolecules {\bf 29}, 7767--7775 (1996).

Ward, I.M. and J.M. Wolfe, 
``The non-linear mechanical behaviour of polypropylene fibers
under complex loading programmes,"
J. Mech. Phys. Solids {\bf 14}, 131--140 (1966).

Wineman, A.S.,
``Branching of strain histories for nonlinear viscoelastic solids
with a strain clock,"
Acta Mech. {\bf 153}, 15--21 (2002).

Wortmann, F.-J. and K.V. Schulz, 
``Non-linear viscoelastic performance of Nomex, 
Kevlar and polypropylene fibres in a single-step 
stress relaxation test: 
1. Experimental data and principles of analysis,"
Polymer {\bf 35}, 2108 (1994).

Wortmann, F.-J. and K.V. Schulz, 
``Non-linear viscoelastic performance of Nomex, Kevlar 
and polypropylene fibres in a single step stress relaxation test:
2. Moduli, viscocities and isochronal stress/strain curves,"
Polymer {\bf 36}, 2363--2369 (1995).

Zhang, C. and I.D. Moore, 
``Nonlinear mechanical response of high density polyethylene.
1: Experimental investigation and model evaluation,"
Polym. Eng. Sci. {\bf 37}, 404--413 (1997).

Zhang, X.C., M.F. Butler, and R.E. Cameron,
``The relationships between morphology, irradiation and
the ductile--brittle transition of isotactic polypropylene,"
Polym. Int. {\bf 48}, 1173--1178 (1999).
\newpage
\section*{List of figures}
\parindent 0 mm

{\bf Figure 1:}
The engineering stress $\sigma$ MPa (unfilled circles)
and the true stress $\sigma_{\rm t}$ MPa (filled circles) versus 
elongation ratio $\lambda$ in a tensile test.
Symbols: experimental data
\vspace*{2 mm}

{\bf Figure 2:}
The strain $\epsilon$ versus time $t$ s in a tensile 
creep test with an engineering stress $\sigma^{0}$ MPa.
Circles: experimental data.
Curve 1: $\sigma_{1}^{0}=10.0$;
curve 2: $\sigma_{2}^{0}=15.0$;
curve 3: $\sigma_{3}^{0}=20.0$;
curve 4: $\sigma_{4}^{0}=25.0$;
curve 5: $\sigma_{5}^{0}=30.0$;
curve 6: $\sigma_{6}^{0}=30.38$.
Solid lines: results of numerical simulation
\vspace*{2 mm}

{\bf Figure 3:}
The Hencky strain $\epsilon_{H}$ versus time $t$ s in a tensile 
creep test with an engineering stress $\sigma^{0}$ MPa.
Symbols: experimental data.
Unfilled circles: $\sigma_{4}^{0}=25.00$;
filled circles: $\sigma_{5}^{0}=30.00$;
asterisks: $\sigma_{6}^{0}=30.38$;
diamonds: $\sigma_{7}^{0}=30.94$;
triangles: $\sigma_{8}^{0}=32.80$.
The lines AA$^{\prime}$ and  BB$^{\prime}$ indicate the strains 
corresponding to transitions from the primary creep 
to the secondary creep and from the secondary creep 
to the ternary creep, respectively
\vspace*{2 mm}

{\bf Figure 4:}
The engineering stress $\sigma$ MPa
versus time $t$ s in a tensile relaxation test 
with a longitudinal strain $\epsilon^{0}$.
Circles: experimental data.
Solid lines: results of numerical simulation.
Curve 1: $\epsilon_{1}^{0}=0.05$;
curve 2: $\epsilon_{2}^{0}=0.10$;
curve 3: $\epsilon_{3}^{0}=0.15$;
curve 4: $\epsilon_{4}^{0}=0.20$
\vspace*{2 mm}

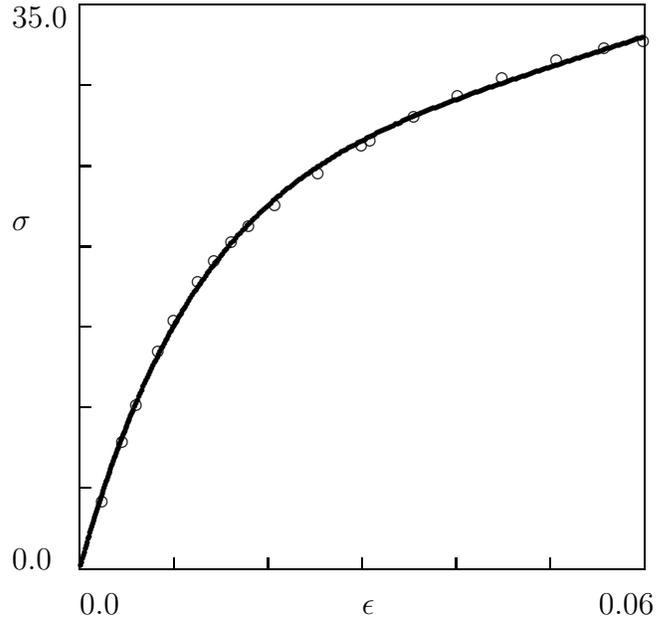
{\bf Figure 5:}
The engineering stress $\sigma$ MPa
versus strain $\epsilon$ in a tensile test.
Circles: experimental data.
Solid line: results of numerical simulation
\vspace*{2 mm}

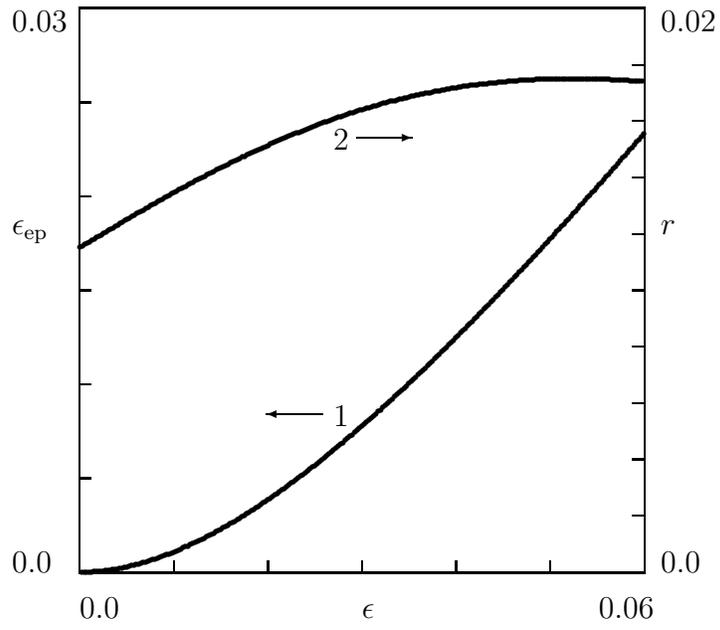
{\bf Figure 6:}
The elastoplastic strain $\epsilon_{\rm ep}$ (curve 1)
and the ratio $r$ of the rate of elastoplastic strain 
to the engineering stress (curve 2) versus strain $\epsilon$ 
in a tensile test.
Solid lines: results of numerical simulation
\vspace*{2 mm}

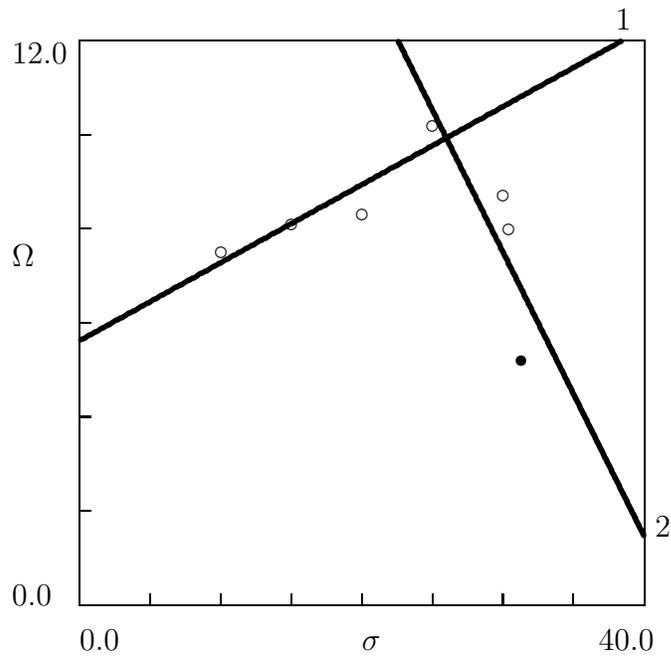
{\bf Figure 7:}
The average potential energy of meso-regions $\Omega$
versus engineering stress $\sigma$ MPa.
Symbols: treatment of observations.
Unfilled circles: creep tests;
filled circle: relaxation test.
Solid lines: approximation of the experimental data by Eq. (33).
Curve 1: $\Omega_{0}=5.62$, $\Omega_{1}=0.17$;
curve 2: $\Omega_{0}=25.66$, $\Omega_{1}=-0.60$
\vspace*{2 mm}

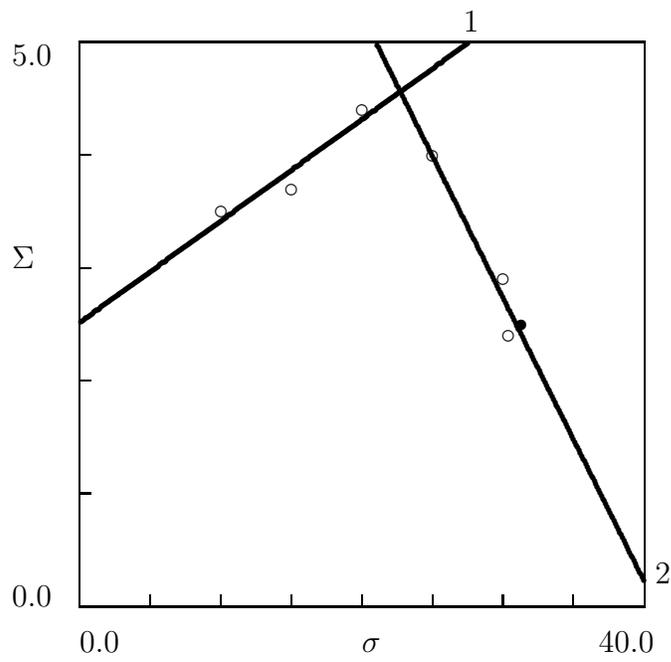
{\bf Figure 8:}
The standard deviation of potential energies of meso-regions $\Sigma$
versus engineering stress $\sigma$ MPa. 
Symbols: treatment of observations.
Unfilled circles: creep tests;
filled circle: relaxation test.
Solid lines: approximation of the experimental data by Eq. (33).
Curve 1: $\Sigma_{0}=2.52$, $\Sigma_{1}=0.09$;
curve 2: $\Sigma_{0}=10.28$, $\Sigma_{1}=-0.26$
\vspace*{2 mm}

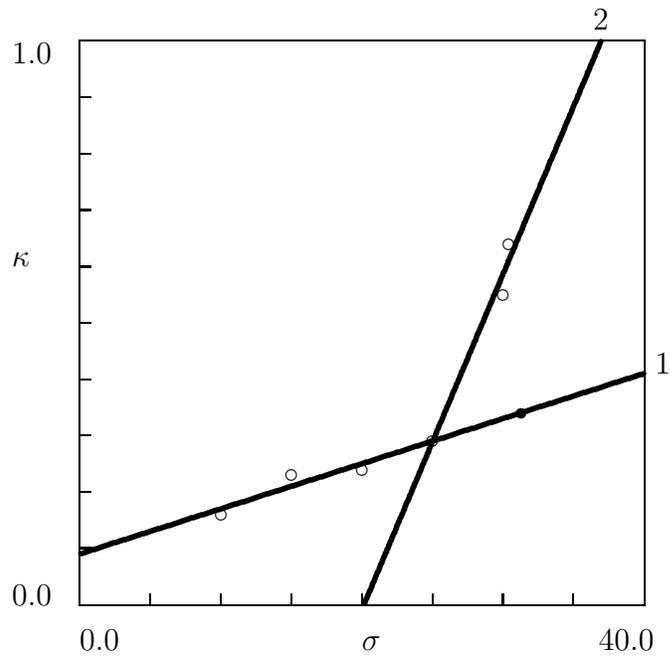
{\bf Figure 9:}
The concentration of active meso-regions $\kappa$
versus engineering stress $\sigma$ MPa.
Symbols: treatment of observations.
Unfilled circles: creep tests;
filled circle: relaxation test.
Solid lines: approximation of the experimental data by Eq. (33).
Curve 1: $\kappa_{0}=0.0900$, $\kappa_{1}=0.0080$;
curve 2: $\kappa_{0}=-1.1995$, $\kappa_{1}=0.0595$
\vspace*{2 mm}

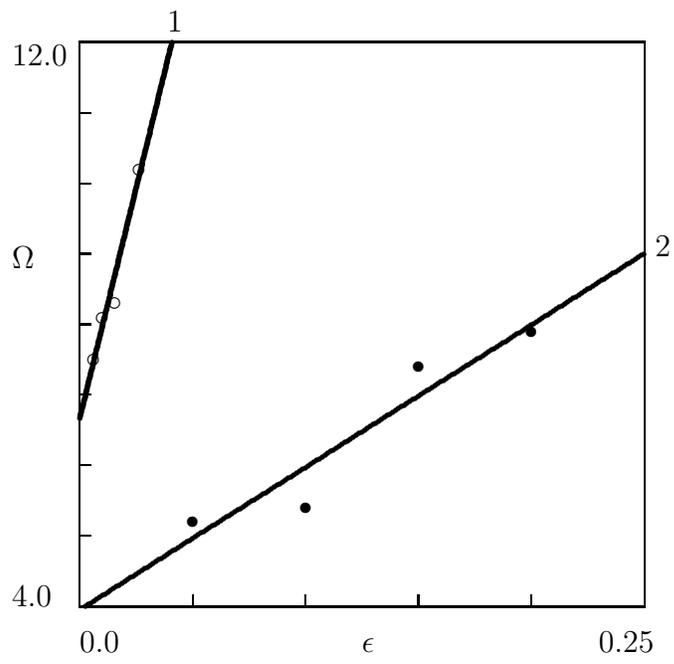
{\bf Figure 10:}
The average potential energy of meso-regions $\Omega$
versus longitudinal strain $\epsilon$.
Symbols: treatment of observations.
Unfilled circles: creep tests;
filled circles: relaxation tests.
Solid lines: approximation of the experimental data by Eq. (35).
Curve 1: $\Omega_{0}=6.66$, $\Omega_{1}=130.66$;
curve 2: $\Omega_{0}=3.95$, $\Omega_{1}=20.20$
\vspace*{2 mm}

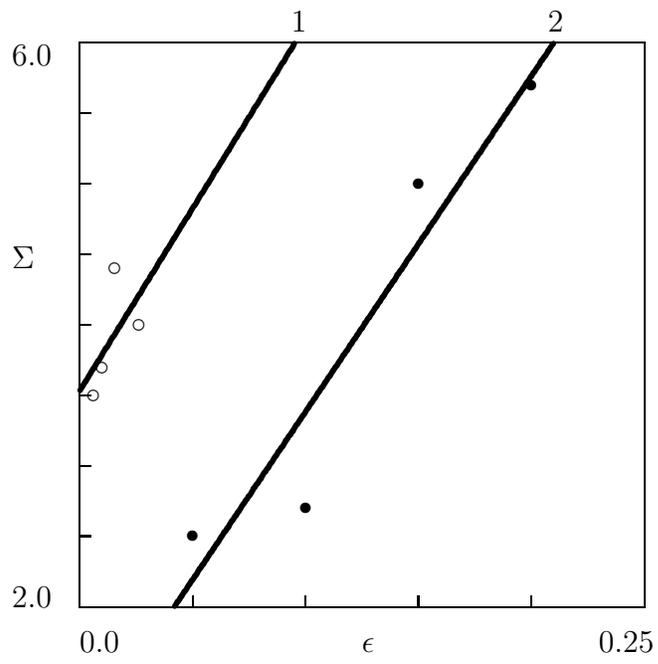
{\bf Figure 11:}
The standard deviation of potential energies of meso-regions $\Sigma$
versus longitudinal strain $\epsilon$.
Symbols: treatment of observations.
Unfilled circles: creep tests;
filled circles: relaxation tests.
Solid lines: approximation of the experimental data by Eq. (35).
Curve 1: $\Sigma_{0}=3.53$, $\Sigma_{1}=25.88$;
curve 2: $\Sigma_{0}=1.00$, $\Sigma_{1}=23.80$
\vspace*{2 mm}

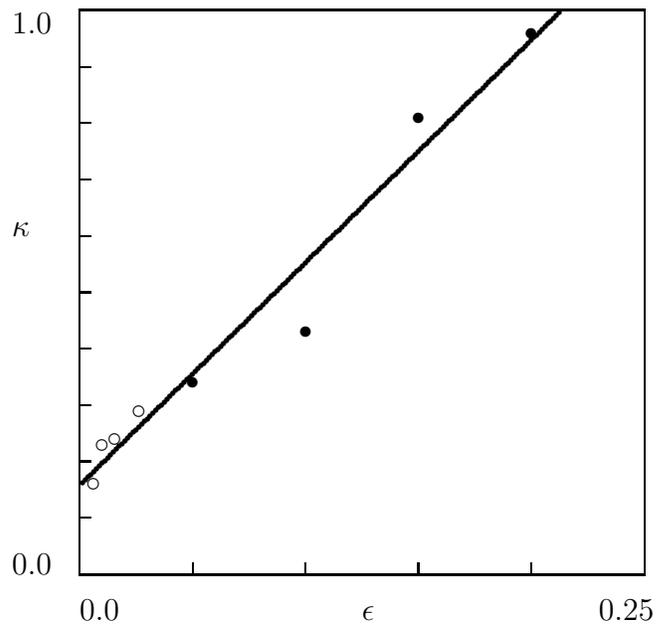
{\bf Figure 12:}
The concentration of active meso-regions $\kappa$
versus longitudinal strain $\epsilon$.
Symbols: treatment of observations.
Unfilled circles: creep tests;
filled circles: relaxation tests.
Solid line: approximation of the experimental data by Eq. (35)
with $\kappa_{0}=0.16$, $\kappa_{1}=3.95$
\vspace*{150 mm}

\setlength{\unitlength}{0.75 mm}
\begin{figure}[tbh]
\begin{center}

\end{center}
\vspace*{10 mm}

\caption{}
\end{figure}

\begin{figure}[tbh]
\begin{center}
%
\end{center}
\vspace*{10 mm}

\caption{}
\end{figure}

\begin{figure}[tbh]
\begin{center}
%
\end{center}
\vspace*{10 mm}

\caption{}
\end{figure}

\begin{figure}[tbh]
\begin{center}
%
\end{center}
\vspace*{10 mm}

\caption{}
\end{figure}

\end{document}